\documentclass[prb,superscriptaddress,amsfonts,twocolumn,showpacs,floatfix]{revtex4}
\usepackage{amsmath}
\usepackage{amsfonts}
\usepackage{amssymb}
\usepackage{bm}
\usepackage{tabularx}
\usepackage{multirow}
\usepackage[dvips]{graphicx,color}

\begin{document}
\title{%
Comparative studies of low temperature X-ray diffraction experiments on $R_{2}$Ir$_{2}$O$_{7}$ ($R=$ Nd, Eu, and Pr)
}

\author{Hiroshi Takatsu}
\affiliation{Department of Physics, Tokyo Metropolitan University, Hachioji-shi, Tokyo 192-0397, Japan}

\author{Kunihiko Watanabe}
\affiliation{Department of Physics, Tokyo Metropolitan University, Hachioji-shi, Tokyo 192-0397, Japan}

\author{Kazuki Goto}
\affiliation{Department of Physics, Tokyo Metropolitan University, Hachioji-shi, Tokyo 192-0397, Japan}

\author{Hiroaki Kadowaki}
\affiliation{Department of Physics, Tokyo Metropolitan University, Hachioji-shi, Tokyo 192-0397, Japan}

\date{\today}

\begin{abstract}
The cubic symmetry of pyrochlore iridium oxides $R_{2}$Ir$_{2}$O$_{7}$ ($R=$ Nd, Eu, and Pr) has 
been investigated by high resolution X-ray diffraction experiments down to 4~K,
in order to clarify the relationship between the metal-insulator transition (MIT) and 
the small structural phase transition suggested by Raman scattering experiments in these compounds.
We have found that a small negative thermal expansion of the order of $10^{-3}$~\AA~
appears only in Nd$_2$Ir$_{2}$O$_{7}$ below the MIT, $T_{\rm MIT}=34$~K, 
ascribable to the magnetovolume effect of the long-range order of Ir moments.
However, 
any breaking of the cubic symmetry of three iridates has not been observed as appearance of
superlattice reflections nor splittings of cubic reflections below $T_{\rm MIT}$.
These results imply that 
lowering of the cubic symmetry plays a minor role for the change in the electronic state of
these compounds, while a magnetic order of Ir moments plays a major role for the MIT.
\end{abstract}


\maketitle

\section{Introduction}
Geometrically frustrated magnets with metallic conduction have attracted much attention
because of the realization of 
a new type of electronic and magnetic behavior, originating from the interplay 
between frustrated spins and conduction electrons~\cite{C.Lacroix,Boldrin2012}. 
The pyrochlore iridates $R_2$Ir$_2$O$_7$ ($R=$ Y, Pr--Lu) are one of the candidates to study such a correlated state,
where an unconventional anomalous Hall effect~\cite{Machida2007,Y.MachidaNature2009}, 
giant magnetoresistance~\cite{MatsuhiraJPSJ2013,DisselerPRB2013}, heavy Fermion behavior~\cite{NakatsujiPRL2006,YanagishimaJPJS2001}, 
and metal-insulator transition (MIT)~\cite{YanagishimaJPJS2001,MatsuhiraJPSJ2007,MatsuhiraJPSJ2011,IshikawaPRB2012,UedaPRL2012} have been observed.
Band calculations~\cite{KooJSSC1998} indicate the importance of Ir 5$d$ electrons
which contribute to novel metallic properties of these compounds:
the conduction band near the Fermi level consists of Ir $5d$ and O $2p$ orbitals in a metallic state,
and
the formal valence of Ir ions is expected to be tetravalent (Ir$^{4+}$, $5d$ $t_{2g}^5$)~\cite{FukazawaJPSJ2002}.
More recent theoretical considerations~\cite{PesinNP2010} indicate the effective 
total-angular-momentum $J_{\rm eff}=1/2$ state of Ir$^{4+}$ 
and the significance of the strong spin orbit interaction and electron correlation inherent to the Ir element,
which can lead to exotic electronic and magnetic phases, 
including a topological Mott insulator~\cite{PesinNP2010,YangPRB2010,KargarianPRB2011,KuritaJPSJ2011,KrempaPRB2010,GuoPRL2009},
a Weyl semimetal~\cite{WanPRB2011,KrempaPRB2012,G.ChenPRL2012}, and an axion insulator~\cite{WanPRB2011,G.ChenPRL2012,A.GoPRL2012}.

Experimentally, most of $R_2$Ir$_2$O$_7$ compounds exhibit MIT 
in accordance with magnetic phase transitions~\cite{YanagishimaJPJS2001,MatsuhiraJPSJ2007}. 
Recent studies on systematic replacement of $R$ ions~\cite{MatsuhiraJPSJ2011} 
indicated that the transition temperature of the MIT, 
$T_{\rm MIT}$, decreases with increasing the ion radius of rare earth ions. 
Interestingly, it is found that
the system becomes metallic in the vicinity of $R=$ Nd and Pr,
where the long-range magnetic order is also suppressed~\cite{NakatsujiPRL2006}
and the spin-liquid like phase is suggested~\cite{Y.MachidaNature2009,S.Onoda2010PRL}.
Mechanical pressure experiments~\cite{SakataPRB2011,TaftiPRB2012} 
also indicated an anomalous metallic phase which can appear in the suppression of the MIT.

Stimulating these experimental results and 
recent extensive attention to an exotic electronic phase for
$5d$ transition metal systems~\cite{PesinNP2010,KimScience2009,ShitadePRL2009,KrempaARCMP2014}, 
several experimental investigations on $R_2$Ir$_2$O$_7$ compounds 
have also been performed to clarify the origin of the MIT.
Raman scattering experiments~\cite{HasegawaJPCS2010} focused on a relationship between the MIT and 
the crystal-structure phase transition which has been often observed in $3d$ transition metal oxides~\cite{ImadaRMP1998}.
The results for sintered powder samples of $R=$ Sm, Eu and a single crystal of Nd showed that
an additional development of a Raman peak, a signature of the structure phase transition, 
was observed below $T_{\rm MIT}$ for Sm$_2$Ir$_2$O$_7$ and Eu$_2$Ir$_2$O$_7$, 
while no clear signals appear for Nd$_2$Ir$_2$O$_7$.
More recent resonant X-ray diffraction (XRD) experiments~\cite{SagayamaPRB2013} on a single crystal of Eu$_2$Ir$_2$O$_7$,
however, did not observe the crystal-structure phase transition, 
although the experiments clarified  the long-range order of Ir$^{4+}$ moments 
consisting of an all-in all-out (AIAO) magnetic structure. 
Neutron diffraction experiments also suggested the long-range order of 
Ir moments in Nd$_2$Ir$_2$O$_7$~\cite{WatahikiJPCS2011}, where
the AIAO structure of the ordered moments was also expected~\cite{TomiyasuJPSJ2012}.
Therefore, in the magnetic point of view, some electronic correlation effects
relating to the magnetic order of Ir moments are essential for 
the MIT of $R_2$Ir$_2$O$_7$, as in the case of Cd$_2$Os$_2$O$_7$~\cite{YamauraPRL2012}. 
These are purely electronic mechanisms for the emergence of the MIT,
which do not require the crystal-structure phase transition and its effect to the band structure.
A recent theory has also suggested an electronic mechanism,
examining a finite temperature effect for physical properties~\cite{KrempaPRB2013}.
However, the situation is still complicated 
because of the discrepancy of experimental results for the crystal structure~\cite{HasegawaJPCS2010,SagayamaPRB2013},
and of the possible effect of the structural change with the lattice distortion~\cite{YangPRB2010}. 
It is thus essential to clarify the low-temperature crystal structure 
as well as the possible emergence of the crystal-structure phase transition in these compounds.
%

In this study, we have performed low-temperature and high-resolution X-ray diffraction experiments on powder samples of 
Nd$_{2}$Ir$_{2}$O$_{7}$ (NIO),  Eu$_{2}$Ir$_{2}$O$_{7}$ (EIO), and Pr$_{2}$Ir$_{2}$O$_{7}$ (PIO). 
We focused on the previous experimental results of Raman scattering
and expected that the change in the crystal-structure phase transition is very small even though there exists the phase transition. 
We thus used a high resolution X-ray diffractometer which
can detect a  change of the lattice parameter of the order of $10^{-4}$~\AA~\cite{GotoJPSJ2011,note_accuracy_experiments}.
We have measured temperature dependence of the lattice parameter down to 4~K, 
since the temperature change in the lattice parameter can be precisely detected by our setup~\cite{note_accuracy_experiments}.
It is also a reason that if the crystal-structure phase transition occurs,
a related structural change or a precursor of the phase transition are known to be observed in 
the $T$-dependence of lattice parameters, strain, and elastic constants~\cite{Kazei1998,HiranoPRB2003,KimuraJPSJ2009,R.L.Melcher}:
thus in addition to searching the emergence of a superlattice reflection and
splitting of peaks in X-ray diffraction patterns, 
searching of an anomaly in the $T$-dependence of the lattice parameter is also a simple and direct 
way for the clarification of small changes in the crystal-structure phase transition.
We have found a negative thermal expansion that appears only for NIO
at temperatures below $T_{\rm MIT}=34$~K.
The crystal structure retains the cubic pyrochlore structure (space group $Fd\bar{3}m$, No.227)
for all three compounds down to 4~K.  
These results suggest that the crystal-structure phase transition from the cubic pyrochlore structure
is not the main origin of the MIT.
Instead, the magnetic order of Ir$^{4+}$ moments 
could lead to a crucial effect for the emergence of the MIT of pyrochlore iridates, 
which should also affect the magnetovolume effect of these compounds.

\section{Experimental}
Polycrystalline samples of NIO, EIO and PIO were prepared by a standard solid-state reaction~\cite{MatsuhiraJPSJ2007,MatsuhiraJPSJ2011,KimuraJPCS2011}.
The appropriate amounts of 
Nd$_2$O$_3$ (99.99\%, Rare Metallic Co., Ltd.), 
Eu$_2$O$_3$ (99.99\%, Rare Metallic Co., Ltd.) or Pr$_6$O$_{11}$ (99.99\%, Rare Metallic Co., Ltd.)
and IrO$_2$  ($>99.9$\%, Tanaka Kikinzoku Kogyo K. K.) were mixed for 30--60 minutes and pressed into pellets.
The pellets were wrapped with Pt foils and placed in evacuated quartz tubes.
These products were then heated at 1173~K for about 3~days. 
After this reaction, additional 3\% of IrO$_2$ was added.
The pellets of the mixtures, wrapped with Pt foils, 
were heated again at 1473~K for about 10~days in evacuated quartz tubes.
In this process, several intermediate grindings were performed in order to react the samples well.

X-ray powder-diffraction experiments were carried out using a Rigaku SmartLab
powder diffractometer equipped with a Cu K$\alpha_1$ monochromator.
The sample was mounted in a closed-cycle He-gas refrigerator and 
the temperature was controlled from 300 to 4~K. 
Details of the confirmation of the sample quality are described in a later section.

To check physical properties of obtained powder samples, 
the specific heat ($C_{P}$) and dc magnetic susceptibility ($M/H$) were measured with 
a commercial calorimeter (Quantum Design, PPMS) and with 
a SQUID magnetometer (Quantum Design, MPMS).
The electrical resistivity $\rho$ was also measured by using a standard four-probe method
for rectangular samples cut out from pellets.

\section{Results}
\label{result}
\begin{figure}[t]
\begin{center}
 \includegraphics[width=0.45\textwidth]{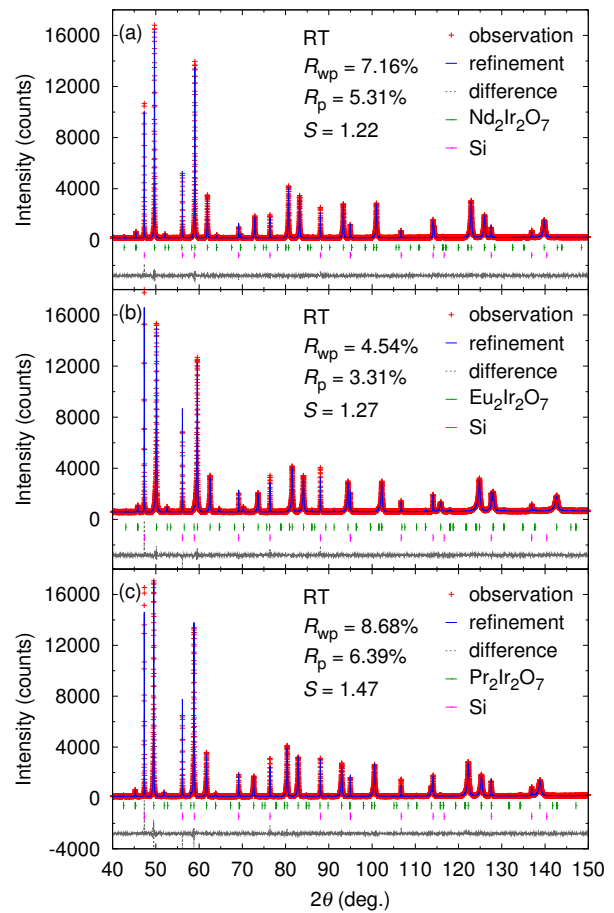}
\caption{
(Color online)
X-ray diffraction patterns of (a) Nd$_{2}$Ir$_{2}$O$_{7}$, (b) Eu$_{2}$Ir$_{2}$O$_{7}$, and 
(c) Pr$_{2}$Ir$_{2}$O$_{7}$ measured at room temperature (RT). 
Observed and refined data are shown by crosses and solid curves, respectively.
The difference between the data and the model is plotted by the dashed curves in the lower part. 
Vertical bars represent positions of the Bragg reflections. 
Si powder was mixed as a reference for the Rietveld analysis 
in order to estimate the lattice constant of the samples precisely.
}
\label{fig.1}
\end{center}
\end{figure}
%
Before going to low temperature XRD experiments,
we have checked the sample quality at room temperature (RT). 
Figure~\ref{fig.1} shows the diffraction patterns and the results of 
Rietveld refinement by using RIETAN-FP~\cite{Izumi2007}.
We confirmed that 
XRD patterns of NIO, EIO, and PIO were reasonably fitted by 
parameters of the cubic pyrochlore structure with the space group $Fd\bar{3}m$.
The refined structure parameters are listed in Table~\ref{table_1}.
The final $R$ factors of the refinements are 
$R_{\rm wp} = 7.16$\%, $R_{\rm e} = 5.88$\%, and $R_{\rm p} = 5.31$\% for NIO, 
$R_{\rm wp} = 4.54$\%, $R_{\rm e} = 3.57$\%, and $R_{\rm p} = 3.31$\% for EIO, and
$R_{\rm wp} = 8.68$\%, $R_{\rm e} = 5.90$\%, and $R_{\rm p} = 6.39$\% for PIO, respectively.
The goodness-of-fit parameter, $S=R_{\rm wp}/R_{\rm e}$, was $S=1.21$, $1.27$, and $1.47$ 
for NIO, EIO, and PIO, respectively, indicating that the qualities of the fitting are good.
Note that 
we confirmed a few amount ($<1$\%) of an impurity phase such as Nd$_{9.33}$(SiO$_4$)$_6$O$_2$
in our samples, however it does not affect the peak profiles and intensities of main peaks such as (440), (444), and (800)
for the low-temperature XRD experiments and physical properties.
We thus considered that our powder samples were reasonable for the experiments.
\begin{table}[t]
\begin{center}
 \caption{Structure parameters at RT for the powder samples of $R_2$Ir$_2$O$_7$ ($R=$ Nd, Eu, and Pr)
 of Fig.~\ref{fig.1}, refined by Rietveld analysis.
 The analysis was performed assuming the space group $Fd\bar{3}m$ (No. 227). 
 The lattice constant $a$ is obtained as $a = 10.3768(2)$~\AA\, for Nd$_2$Ir$_2$O$_7$, $a = 10.2857(3)$~\AA\, for Eu$_2$Ir$_2$O$_7$,
 and $a = 10.4105(3)$~\AA\, for Pr$_2$Ir$_2$O$_7$, respectively. 
 The $U_{\rm iso}$ parameter of 16c(Ir) of Nd$_2$Ir$_2$O$_7$ and Pr$_2$Ir$_2$O$_7$ was fixed in the analysis.
 $U_{\rm iso}$ values of 48f(O) and 8b(O') of three compounds were also fixed in the analysis.
 }
 \begin{tabular*}{0.48\textwidth}{@{\extracolsep{\fill}}cccccl}
  \hline\hline
   Atom    & Site   & x        & y   & z     & {$U_{\mathrm{iso}}$ ($10^{-3}$\AA$^2$)}           \rule{0mm}{4mm}   \\ \hline
   \,Nd    & 16d    & 1/2      & 1/2 & 1/2   & \quad\,\,\, 1.3(1)               \rule{0mm}{4mm}   \\
   \,Ir    & 16c    & 0        & 0   & 0     & \quad\,\,\, 1.7(fix)            \rule{0mm}{3.5mm} \\
   \,O     & 48f    & 0.330(1) & 1/8 & 1/8   & \quad\,\,\, 5 \,\,\,(fix)            \rule{0mm}{3.5mm} \\
   \,O'    & 8b     & 3/8      & 3/8 & 3/8   & \quad\,\,\, 5 \,\,\,(fix)            \rule{0mm}{3.5mm} \\ \hline
   \,Eu    & 16d    & 1/2      & 1/2 & 1/2   & \quad\,\,\, 4.2(2)               \rule{0mm}{4mm}   \\
   \,Ir    & 16c    & 0        & 0   & 0     & \quad\,\,\, 1.7(1)               \rule{0mm}{3.5mm} \\
   \,O     & 48f    & 0.336(1) & 1/8 & 1/8   & \quad\,\,\, 5 \,\,\,(fix)            \rule{0mm}{3.5mm} \\
   \,O'    & 8b     & 3/8      & 3/8 & 3/8   & \quad\,\,\, 5 \,\,\,(fix)            \rule{0mm}{3.5mm} \\ \hline
   \,Pr    & 16d    & 1/2      & 1/2 & 1/2   & \quad\,\,\, 1.9(1)               \rule{0mm}{4mm}   \\
   \,Ir    & 16c    & 0        & 0   & 0     & \quad\,\,\, 1.7(fix)            \rule{0mm}{3.5mm} \\
   \,O     & 48f    & 0.330(1) & 1/8 & 1/8   & \quad\,\,\, 5 \,\,\,(fix)            \rule{0mm}{3.5mm} \\
   \,O'    & 8b     & 3/8      & 3/8 & 3/8   & \quad\,\,\, 5 \,\,\,(fix)            \rule{0mm}{3.5mm} \\
  \hline\hline
 \end{tabular*}
 \label{table_1}
\end{center}
\end{table}

\begin{figure}[t]
\begin{center}
 \includegraphics[width=0.45\textwidth]{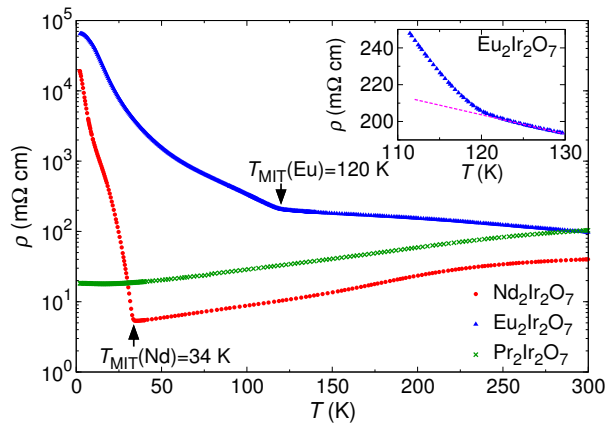}
\caption{
(Color online)
Temperature dependence of $\rho$ of polycrystalline samples of 
$R_2$Ir$_2$O$_7$ ($R=$ Nd, Eu, and Pr). 
Inset shows $\rho$ of Eu$_2$Ir$_2$O$_7$ around $T_{\rm MIT}=120$~K.
The dashed line in the inset is a guide to the eyes for the slope above $T_{\rm MIT}$.
}
\label{fig.2}
\end{center}
\end{figure}

\begin{figure*}[t]
\begin{center}
 \includegraphics[width=0.90\textwidth]{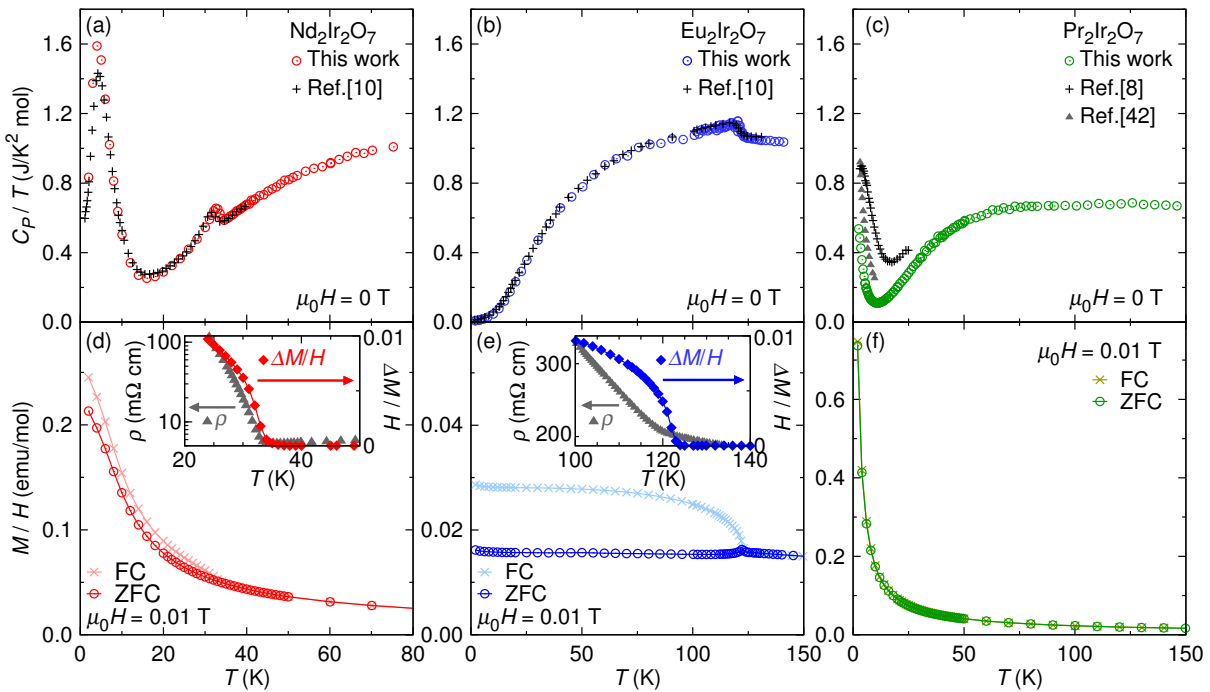}
\caption{
(Color online)
Temperature dependences of (a)--(c) $C_P/T$ and (d)--(f) $M/H$ for powder samples of Nd$_2$Ir$_2$O$_7$, Eu$_2$Ir$_2$O$_7$, 
and Pr$_2$Ir$_2$O$_7$, respectively.
The samples used in this study show qualitatively the same behaviors of previous resports~\cite{MatsuhiraJPSJ2011,YanagishimaJPJS2001,TokiwaNM2014}.
For Pr$_2$Ir$_2$O$_7$, there exist quantitative differences in the upturn of $C_P/T$ at low temperatures,
which may be attributable to the small difference of the off-stoichiometry of samples.
A similar behavior has been observed in the systematic change of $x$ for
the zirconium analogs of Pr$_{2+x}$Zr$_{2-x}$O$_{7+y}$ ($-0.02\le x \le0.02$)~\cite{KoohpayehJCG2014}.
Figures (d)--(f) include results of both zero-field-cooling (ZFC) and field-cooling (FC) measurements of $M/H$.
Insets of (d) and (e) show comparisons between $\rho$ (for a log scale) and $\varDelta M/H$ for Nd$_2$Ir$_2$O$_7$ and Eu$_2$Ir$_2$O$_7$, 
respectively. Here $\varDelta M/H$ is a difference of the susceptibility between the ZFC measurement ($M_{\rm ZFC}/H$) 
and the FC measurement ($M_{\rm FC}/H$): $\varDelta M/H = M_{\rm FC}/H - M_{\rm ZFC}/H$.
}
\label{fig.3}
\end{center}
\end{figure*}

The prepared samples exhibit qualitatively the same behaviors of $M/H$, $C_{P}$, and $\rho$ as those of 
the previous reports (Figs.~\ref{fig.2} and \ref{fig.3})~\cite{MatsuhiraJPSJ2007,MatsuhiraJPSJ2011,KimuraJPCS2011,IshikawaPRB2012,TokiwaNM2014}. 
We confirmed that the second-order phase transition of the MIT appear at $T_{\rm MIT}=34$~K for NIO and at $T_{\rm MIT}=120$~K for EIO (Fig.~\ref{fig.2}).
For PIO, the metallic behavior was confirmed down to 2~K of the present lowest measured temperature,
where PIO does not show any signatures of a magnetic phase transition and MIT.
It can thus be considered that PIO becomes a reference compound for NIO and EIO when comparing behaviors with and without the MIT.
Note that the absolute value of $\rho$ of PIO is slightly larger than that of Ref.~[10], 
while it is almost the same as that of Ref.~[33]. 
This result is probably due to the grain boundaries in the pelletized samples,
which brings about a weak localization of conduction electrons.
It is also noted here that a slight discrepancy 
between our sample and previous polycrystalline samples~\cite{YanagishimaJPJS2001,MatsuhiraJPSJ2011} 
was observed in the $T$-dependence of $\rho$ for EIO below $T_{\rm MIT}$. 
This result may come from the effect of a small amount of the off-stoichiometry of samples~\cite{IshikawaPRB2012}.
We will discuss it in a later section.

\begin{figure}[t]
\begin{center}
 \includegraphics[width=0.45\textwidth]{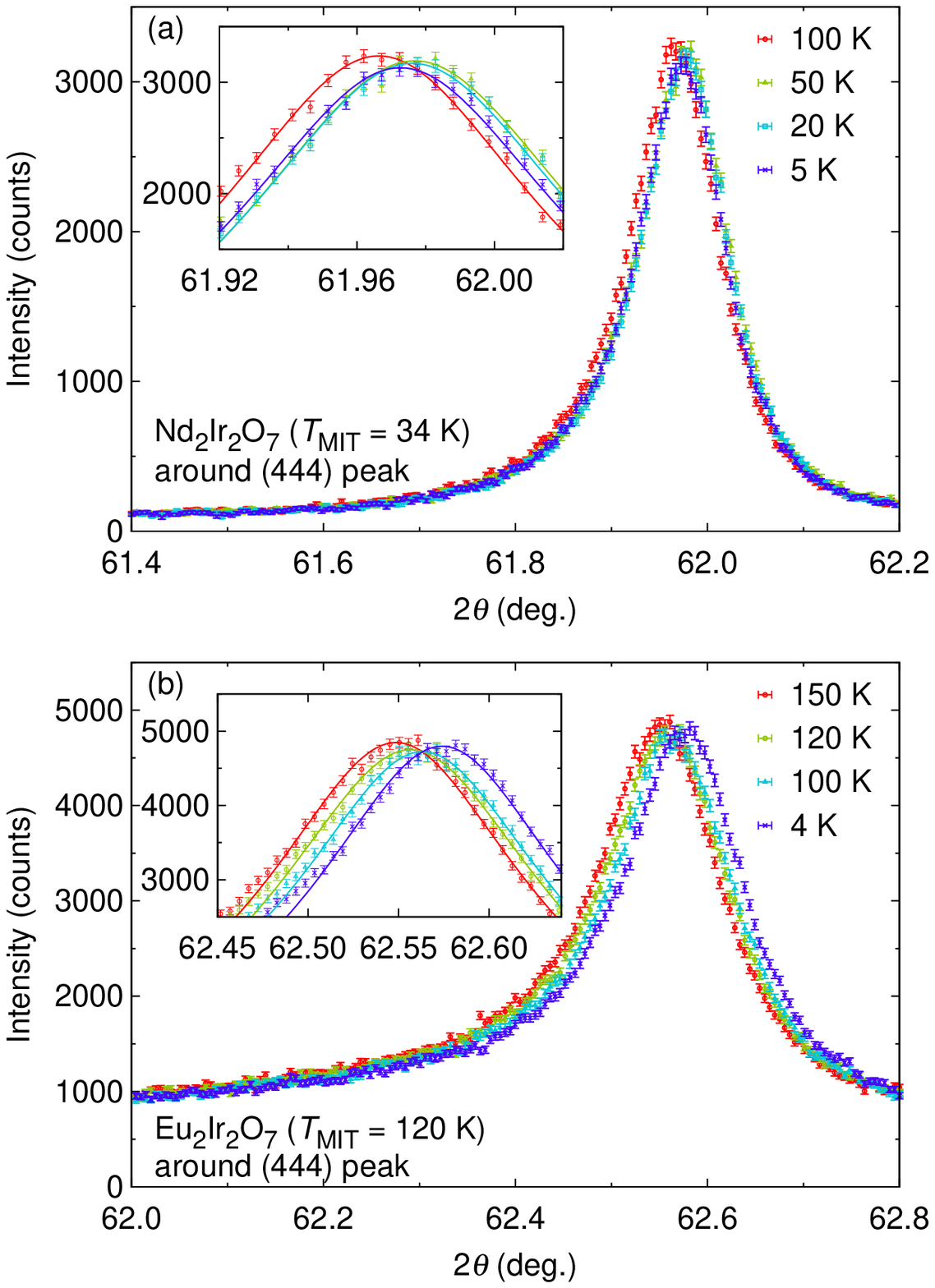}
\caption{
(Color online)
2$\theta$-$\theta$ scans around the (444) reflections for (a) Nd$_2$Ir$_2$O$_7$ and for (b) Eu$_2$Ir$_2$O$_7$
at temperatures above and blow $T_{\rm MIT}$.
Insets show enlargements around the peak positions.
}
\label{fig.4}
\end{center}
\end{figure}
\begin{figure}[t]
\begin{center}
 \includegraphics[width=0.45\textwidth]{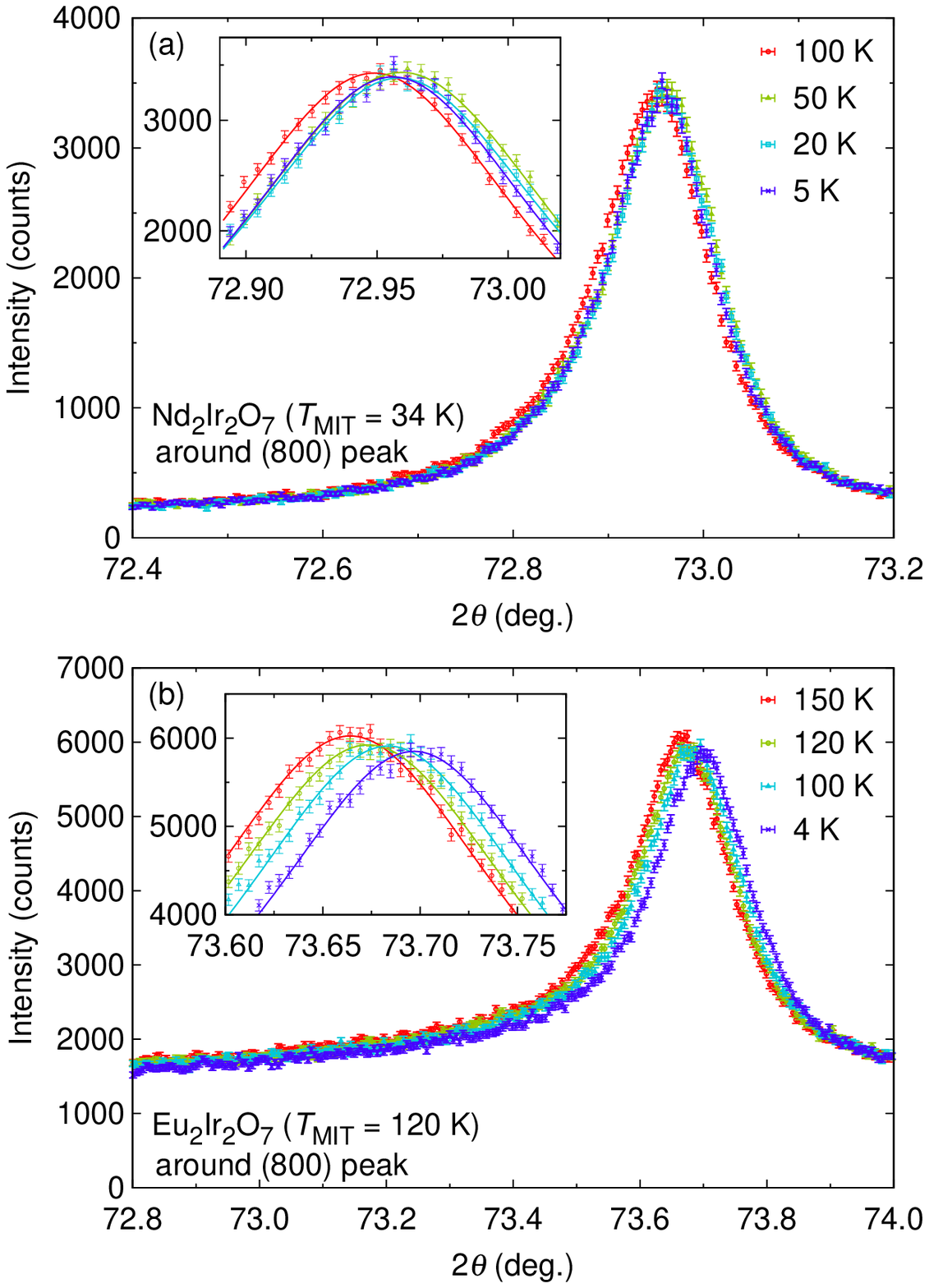}
\caption{
(Color online)
2$\theta$-$\theta$ scans around the (800) reflections for (a) Nd$_2$Ir$_2$O$_7$ and for (b) Eu$_2$Ir$_2$O$_7$
at temperatures above and blow $T_{\rm MIT}$.
Insets show enlargements around the peak positions.
}
\label{fig.5}
\end{center}
\end{figure}
Next, we have checked the possibility of a crystal-structure phase transition for NIO and EIO
by searching the emergence of a superlattice reflection and splitting of peaks in the XRD patterns at low temperatures: 
these are experimental signatures of a crystal-structure phase transition. 
However, we did not observe any superlattice reflections
nor splitting and broadening of certain peaks at temperatures below $T_{\rm MIT}$: 
i.e., $T_{\rm MIT} = 34$~K for NIO and $T_{\rm MIT} = 120$~K for EIO.
Even though there exists a superlattice reflection,
peak intensities are expected to be below 0.01\% of the maximum peak counts of the (111) peak,
where $2\times10^5\sim1\times10^6$ counts were integrated in our experiments.
It is thus considered that the crystal symmetry retains the cubic pyrochlore structure 
even for a low temperature phase of NIO and of EIO.
In Figs.~\ref{fig.4} and \ref{fig.5}, we show representative peak profiles around the (444) and (800) peaks of NIO and EIO, respectively.
One can see no splitting and broadening of the peaks below $T_{\rm MIT}$. 
In these experiments, the peak positions are determined within a very small experimental error of $0.001^\circ$,
as in the case of previous experiments on the powder sample of Tb$_2$Ti$_2$O$_7$~\cite{GotoJPSJ2011}.

\begin{figure}[t]
\begin{center}
 \includegraphics[width=0.45\textwidth]{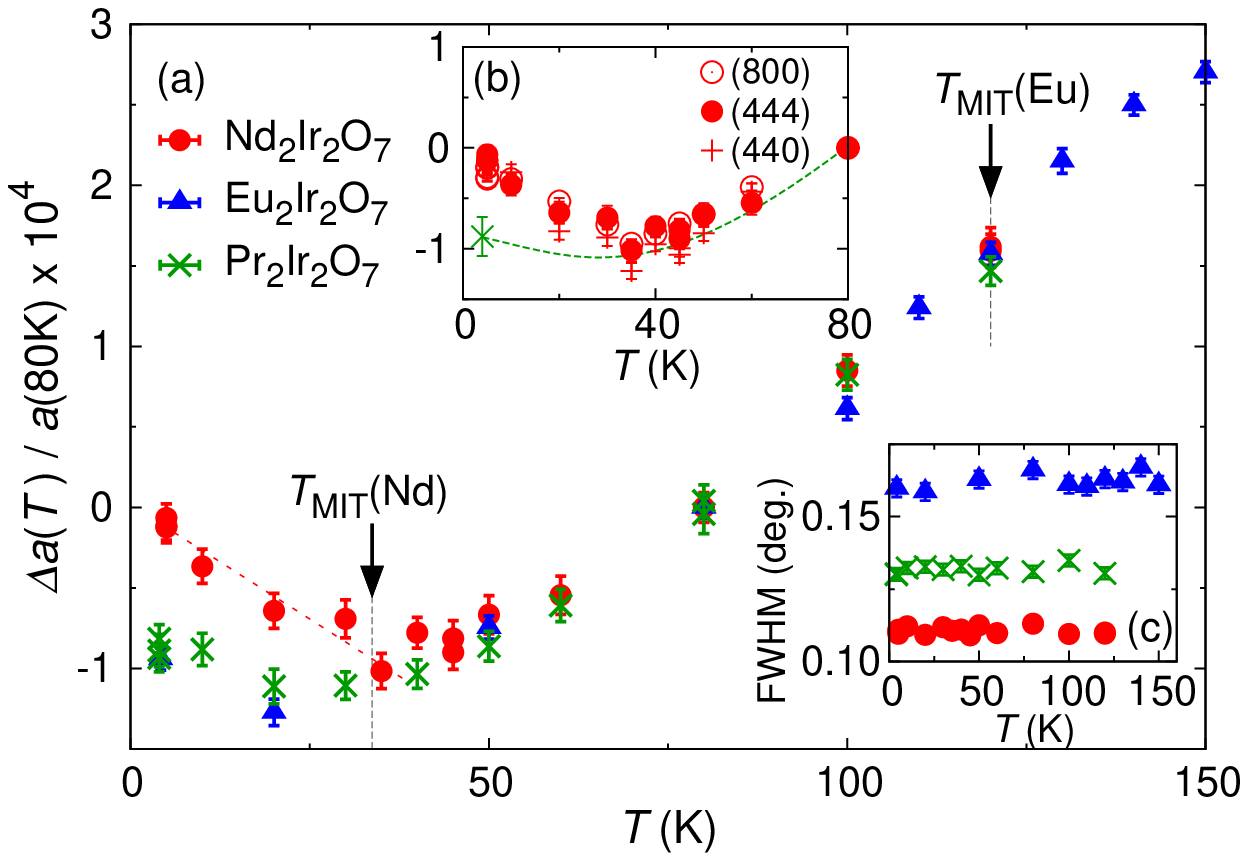}
\caption{
(Color online)
(a) Temperature dependence of the lattice constant $a(T)$ of $R_2$Ir$_2$O$_7$ ($R=$ Nd, Eu, and Pr), 
which were estimated from the peak position of the (444) reflections. 
The data was normalized by the data at $T=80$~K: $\varDelta a(T)/a(80~{\rm K}) = a(T)/a(80~{\rm K})-1$.
The dashed line for Nd$_2$Ir$_2$O$_7$ below $T_{\rm MIT}$ indicates the linear fit result corresponding to 
the estimation of $\alpha = -1.5(1)\times10^{-6}$~K$^{-1}$.
(b) $\varDelta a(T)/a(80~{\rm K})$ of Nd$_2$Ir$_2$O$_7$ around $T_{\rm MIT} = 34$~K.
The vertical axis is the same as that of (a). 
In this plot, $a(T)$ is examined from peak positions at the (800), (444), and (440) peaks.
The dashed line is extrapolated from the data of Pr$_2$Ir$_2$O$_7$ in (a).
(c) Temperature dependence of the full width at half maximum (FWHM) of the (444) peak of three compounds.
The definition of the symbols is the same as that in (a).
}
\label{fig.6}
\end{center}
\end{figure}

We found a different temperature dependence of the lattice parameter for EIO and NIO. 
EIO shows a positive thermal expansion, while NIO shows a negative thermal expansion on warming. 
It is slight but experimentally clear for the peak shift of those compounds 
[insets of Figs.~\ref{fig.4}(a) and (b) and of Figs.~\ref{fig.5}(a) and (b)],
where the peak of EIO and of NIO goes forward and backward around $T_{\rm MIT}$, respectively.
In Fig.~\ref{fig.6}, 
we show the temperature dependence of the lattice parameter $a(T)$ of NIO and EIO, 
estimated from the peak position of the (444) peak at each temperature.
The data was normalized by one at $T=80$~K, $\varDelta a(T)/a(80~{\rm K})$, where $\varDelta a(T) = a(T)-a(80~{\rm K})$.
For reference, we also show the results of PIO in the same experimental way.
One can see that $\varDelta a(T)/a(80~{\rm K})$ continuously decreases on cooling for three compounds.
However, only NIO exhibits the negative thermal expansion below about $T_{\rm MIT} = 34$~K.
The same tendency was also confirmed by using other experimental results for 
converting $a(T)$ from peak positions of e.g. the (800) and (440) peaks for NIO [Fig.~\ref{fig.6}(b)]. 
The linear coefficient of the thermal expansion $\alpha = (1/a(T))(\partial a(T)/\partial T)$
is obtained as $\alpha = -1.5(1)\times10^{-6}$~K$^{-1}$ at $T=35$~K 
and $\alpha =  1.7(1)\times10^{-6}$~K$^{-1}$ at $T=40$~K for NIO.
We can thus estimate the discontinuity of $\alpha$ around $T_{\rm MIT}$ as
$\varDelta \alpha = -3.2(1)\times10^{-6}$~K$^{-1}$.
Note that a very small upturn is also found at temperatures below 20~K for EIO and PIO,
however the value of the changes is one order of magnitude smaller than that for NIO.
In the whole measured temperature range,
the full width at half maximum (FWHM) of profile peaks is almost $T$-independent 
(inset of Fig.~\ref{fig.6} for the (444) peak), 
implying again no reduction of the crystal symmetry below $T_{\rm MIT}$
within the experimental accuracy.

It is worth noting here that we didn't observe any obvious anomaly in $\varDelta a(T)/a(80~{\rm K})$
at about $T=15$~K for NIO, where  
the localized moments of Nd$^{3+}$ are known to exhibit an antiferromagnetic long-range order~\cite{TomiyasuJPSJ2012}.

\section{Discussion}
\label{Dis}
From our low temperature XRD data presented above,
we experimentally revealed that pyrochlore iridates of NIO and EIO
do not show the crystal-structure phase transition at temperatures lower than $T_{\rm MIT}$.
Moreover, within the experimental accuracy, we didn't observe appropriate differences in 
the temperature dependence of the peak positions of (800), (444), and (440) [Fig.~\ref{fig.6}(b)],
which implies the lattice retains the cubic symmetry.
We therefore think that
the crystal-structure phase transition from the cubic pyrochlore structure 
is not the main origin of the MIT of pyrochlore iridates in real materials.
In contrast, our powder samples of NIO and EIO show the MIT and the magnetic anomaly in $M/H$ at almost the same temperature 
[insets of Figs.~\ref{fig.3}(d) and (e)],
implying the long-range order of the Ir$^{4+}$ moments with forming the AIAO magnetic structure~\cite{TomiyasuJPSJ2012,SagayamaPRB2013}.
It is thus considered that the magnetic order with the AIAO magnetic structure of Ir moments 
brings about the change in the electronic state at $T_{\rm MIT}$.
A similar case has been recently pointed out for Cd$_2$Os$_2$O$_7$~\cite{YamauraPRL2012}.
On the basis of the irreducible representation of the AIAO magnetic structure
at the 16c-site atom of the pyrochlore lattice~\cite{YamauraPRL2012,SagayamaPRB2013}, 
the cubic structural symmetry can be preserved at temperatures above and below 
the magnetic phase transition. It is consistent with the present experimental results.

The incompatible behavior of the lattice expansion of NIO and of EIO can be understood semi-quantitatively
by the Ehrenfest relation for a second-order phase transition: i.e.,
$(\partial T_{\rm MIT}/\partial P)_{P\rightarrow0} = T_{\rm MIT}V_{\rm mol}\varDelta\alpha/\varDelta C_{P}$, 
where $V_{\rm mol}$ is the molar volume and  $\varDelta C_{P}$ is
the peak height of the specific heat at $T_{\rm MIT}$.
From present $C_{P}$ and XRD experiments and 
previous pressure-effect experiments~\cite{SakataPRB2011,TaftiPRB2012},
these values are roughly estimated as $(\partial T_{\rm MIT}/\partial P)_{P\rightarrow0} \simeq -10$~K/GPa, 
$\varDelta C_{P}/T_{\rm MIT}\simeq0.06$~J/K$^2$-mol, and $V_{\rm mol}\simeq84$~cm$^3$/mol at $T_{\rm MIT}=34$~K for NIO,
while
$(\partial T_{\rm MIT}/\partial P)_{P\rightarrow0} \simeq -5$~K/GPa, 
$\varDelta C_{P}/T_{\rm MIT}\simeq0.07$~J/K$^2$-mol, and $V_{\rm mol}\simeq82$~cm$^3$/mol at $T_{\rm MIT}=120$~K for EIO,
respectively.
Therefore, the expected value of $\varDelta\alpha$ due to the phase transition can be calculated as
$\varDelta\alpha_{\rm cal} \simeq - 7\times10^{-6}$~K$^{-1}$ for NIO and 
$\varDelta\alpha_{\rm cal} \simeq - 4\times10^{-6}$~K$^{-1}$ for EIO, respectively.
Both compounds are expected to show the negative thermal expansion,
however the change in NIO should appear as a larger effect at $T_{\rm MIT}$.
In our experiments, it was estimated that $\varDelta\alpha = -3.2\times10^{-6}$~K$^{-1}$ for NIO.
The absolute value of the experiment is somewhat smaller than that of the rough estimation above, 
although the order of the value is the same.
We thus think that other extra effects could reduce the absolute value of the negative thermal expansion 
from the expected value for EIO as well.
In any case, in view of the thermodynamic property of a phase transition,
the negative thermal expansion observed in NIO can be understood in terms of
the magnetovolume effect ascribable to the magnetic phase transition of the Ir moments at about $T_{\rm MIT}$.

We finally comment on the discrepancy of the previous results about the crystal structure~\cite{HasegawaJPCS2010,SagayamaPRB2013}
and electrical resistivity~\cite{YanagishimaJPJS2001,MatsuhiraJPSJ2007,MatsuhiraJPSJ2011,IshikawaPRB2012}.
One of the possible origins of these results may be attributed to the sample dependence.
It is known that frustrated magnets often exhibit strong sample dependence due to the frustration among spin interactions and many-body effects.
This tendency has also been found in pyrochlore oxides $A_{2}$$B_{2}$O$_{7}$ 
($A =$ rare earth ions, $B=$ transition metal ions)~\cite{YanagishimaJPJS2001,MatsuhiraJPSJ2007,MatsuhiraJPSJ2011,IshikawaPRB2012,KimuraJPCS2011,H.TakatsuJPCM2012,RossPRB2012,TaniguchiPRB2013,KoohpayehJCG2014}, 
including NIO~\cite{YanagishimaJPJS2001,MatsuhiraJPSJ2011} and 
EIO~\cite{IshikawaPRB2012,YanagishimaJPJS2001,MatsuhiraJPSJ2011}, as well as PIO~\cite{YanagishimaJPJS2001,MatsuhiraJPSJ2011,KimuraJPCS2011}, 
where the distribution of nonstoichiometric concentrations such as $x$ and $y$ of $A_{2+x}$$B_{2-x}$O$_{7+y}$ 
is thought to induce sample dependent physical properties~\cite{IshikawaPRB2012,KimuraJPCS2011,H.TakatsuJPCM2012,RossPRB2012,TaniguchiPRB2013,KoohpayehJCG2014}.
Even in our samples, we also observed a slight discrepancy in EIO for the previous results.
For examples, 
the upturn of $\rho$ of EIO below $T_{\rm MIT}$ 
is suppressed in our samples, as compared with the previous result in Ref.~[10]:
however, it should be noted here that in contrast to both results, 
the sample of EIO in Ref.~[8] retains the metallic behavior without an any signature of the MIT.
Recent studies on single crystalline samples of EIO showed 
that $\rho$ is quite sensitive to the off-stoichiometry $x$ of samples~\cite{IshikawaPRB2012}.
According to this result, $x \simeq 0.02$ is expected for the previous polycrystalline sample of EIO in Ref.~[10], while 
$x \simeq 0.03$ for our samples.
Such a small distribution ($\leq1$\% discrepancy) gives rise to a strong sample dependent physical property in these compounds.
Therefore, we should take care of the sample quality for studying the frustrated pyrochlore magnets,
although this problem could be a signature of exotic materials.
In the present experiments, 
we used identical batch samples for experiments of $C_P$, $M/H$, and $\rho$ as well as the low-$T$ XRD experiments.
We thus consider that present experimental results provide compatible and consistent data for the discussion
about the relationship between the MIT and the structural phase transition in pyrochlore iridates,
i.e., large structural changes do not occur even when the MIT is visible in the materials.

\section{Conclusion}
We have performed the high-resolution and low-temperature X-ray diffraction experiments 
on polycrystalline samples of $R_{2}$Ir$_{2}$O$_{7}$ ($R=$ Nd, Eu, and Pr),
in order to clarify the relationship between the MIT and the crystal-structure phase transition.
We confirmed that the structure symmetry of these compounds is not broken,
preserving the cubic pyrochlore lattice symmetry even below the transition temperature of the MIT:
$T_{\rm MIT} = 34$~K for Nd$_2$Ir$_2$O$_{7}$ and $T_{\rm MIT} = 120$~K for Eu$_2$Ir$_2$O$_{7}$.
We also found the positive thermal expansion for Eu$_2$Ir$_2$O$_{7}$ and Pr$_2$Ir$_2$O$_{7}$,
while the negative thermal expansion for Nd$_2$Ir$_2$O$_{7}$, which may come from the magnetovolume effect 
attributed to the long-range order of Ir magnetic moments.
These results suggest that the crystal-structure phase transition from the cubic pyrochlore structure,
analogously to cases of 3$d$ transition metal oxides,
is not the origin of the MIT of Nd$_2$Ir$_2$O$_{7}$ and Eu$_2$Ir$_2$O$_{7}$.
Instead, the results imply that the magnetic order of Ir moments should 
affect the change in the electronic state of these compounds at $T_{\rm MIT}$.
Experimental studies using high-quality single crystals 
are important for further clarification of the origin of the MIT in pyrochlore iridates.

\section*{Acknowledgment}
This work was partly supported by JSPS KAKENHI Grant Number 24740240.

\bibliography{reference}

\begin{thebibliography}{50}
\bibitem[{not()}]{note_accuracy_experiments}
\bibinfo{note}{{\color[rgb]{0,0,0} In this study, the temperature change in the
  lattice parameter $a$ was detected in a small error of $\varDelta a \sim
  10^{-4}$~\AA\,. This accuracy is realized by the resolution of the absolute
  angle encoder, $1/3600$ degrees, installed in the Rigaku SmartLab powder
  diffractometer used in this study. This means that a peak shift of the
  scattering angle $2\theta$ is very precisely measured by using such
  diffractometer: the temperature change in $a$ can therefore be measured by
  precise observations of the temperature change in $2\theta$ at an appropriate
  peak such as the (444) peak and the (800) peak (e.g., see insets of Figs.~4
  and 5). When converting $\varDelta a \sim 10^{-4}$~\AA\, to the accuracy of
  $2\theta$, it is estimated as $\varDelta2\theta \sim 0.001^{\circ}$, which is
  different from the resolution width of $2\theta$ and is instead supported by
  the resolution of the absolute angle encoder. We also note that the precision
  of $2\theta$ in experiments is down to $0.001^{\circ}$}}.
\end{thebibliography}
\end{document}